\documentclass[a4paper,11pt]{article}
\usepackage{pos}

\newcommand{\nn}{\nonumber}

\newcommand{\beqa}{\begin{eqnarray}}
\newcommand{\eeqa}{\end{eqnarray}}

\title{Lattice study on a tetraquark state $T_{bb}$ in the HAL QCD method}

\author*[a,b]{Sinya Aoki}
\author[c]{Takafumi Aoki}

\affiliation[a]{Center for Gravitational Physics and Quantum Information, Yukawa Institute for Theoretical Physics, Kyoto University,
  Koto 606-8502, Japan}
  
 \affiliation[b]{Interdisciplinary Theoretical and Mathematical Sciences Program (iTHEMS), RIKEN, Wako  351-0198, Japan}
  
 \affiliation[c]{Yukawa Institute for Theoretical Physics, Kyoto University,
  Koto 606-8502, Japan}

\emailAdd{saoki@yukawa.kyoto-u.ac.jp }
\emailAdd{takafumi.aoki@yukawa.kyoto-u.ac.jp}

\abstract{
We investigate a doubly-bottomed tetraquark state $T_{bb}$ $(bb \bar{u}\bar{d})$ with quantum number $I(J^P)=0(1^+)$
in $(2+1)$-flavor lattice QCD.
Using the Non-Relativistic QCD (NRQCD) quark action for $b$ quarks, 
we have extracted the coupled channel potential between $\bar{B}\bar{B}^*$ and $\bar{B}^* \bar{B}^*$ in the  HAL QCD method
at $a \approx 0.09$ {fm} on $32^3\times 64$ lattices.
The potential predicts an existence of a bound  $T_{bb}$ below the $\bar{B}\bar{B}^*$ threshold.
At the physical pion mass $m_\pi\approx140$ {MeV} extrapolated from $m_\pi\approx 410,\, 570,\, 700$ {MeV}, 
 a binding energy with its statistical error is given by  $E_{\rm binding}^{\rm (coupled)} = 83(10)$ MeV from a coupled channel analysis where effects due to virtual $\bar{B}^* \bar{B}^*$ states are included through the coupled channel potential, while
 we obtain $E_{\rm binding}^{\rm (single)} = 155(17)$ MeV only from a potential for a single $\bar{B}\bar{B}^*$ channel. 
This difference indicates  that the effect from virtual $\bar{B}^* \bar{B}^*$ states is sizable to the binding energy of $T_{bb}$. 
Adding $\pm 20$ MeV as empirical systematic error caused by  the NRQCD approximation for $b$ quarks, 
our estimate of the $T_{bb}$ binding energy becomes  $83(10)(20)$ MeV.
}

\FullConference{%
  The 39th International Symposium on Lattice Field Theory (Lattice2022),\\
  8-13 August, 2022 \\
  Bonn, Germany 
}

\begin{document}
\begin{flushright}
YITP-22-148
\end{flushright}
\maketitle

\section{Introduction}
A tetraquark state is one of exotic states  in QCD other than mesons and baryons and is made of two quarks and two antiquarks.
Recently heavy tetraquark states, made of two heavy quarks and two light antiquarks, attract much attentions,   probably because 
they are genuine tetraquark states as $QQ \bar q\bar q$ or $qq \bar Q\bar Q$, where $Q$ and $q$ are a heavy and a light quarks respectively. 
We denote such a heavy tetraquark state as $T_{QQ}$.
Indeed, an observation of  $T_{cc}$ ($cc\bar u\bar d$) state  just below the $D^{* +} D^0$ threshold was reported by the LHCb collaboration\cite{LHCb:2021auc}.
While results from previous lattice QCD calculations\cite{Guerrieri:2014nxa,Ikeda:2013vwa} were inconclusive,
the latest lattice QCD study\cite{Padmanath:2022cvl} suggests a virtual state for $T_{cc}$.
 
Another candidate for  heavy tetraquark states is $T_{bb}$ ($bb \bar{u}\bar{d}$ or $\bar{b}\bar{b} ud$), which has not been observed yet but is more likely to exist as a bound state than $T_{cc}$, since a force between two $\bar b$ is probably described by a screened Coulomb potential, 
generated as a mixture of one-gluon-exchange at short distance and a screening due to light quarks at long distance.  

There are several lattice studies on $T_{bb}$.
The single channel potential between $\bar B$ and $\bar B^*$, calculated using the static quark action to treat $b$ quarks on the lattice,
predicted an existence of one bound state in the $I(J^P)=0(1^+)$ channel, whose binding energy is extrapolated to the physical pion mass as $E_B=90^{+43}_{-36}$\cite{Bicudo:2015kna}. The coupled channel potential  between $\bar B \bar B^*$ and $\bar B^* \bar B^*$ reduces the binding energy  to $E_B=59^{+30}_{-38}$\cite{Bicudo:2016ooe}.
Direct spectrum calculations in lattice QCD using NRQCD for $b$ quarks to include their moving in space
increase the binding energy significantly to $ E_B\simeq 120 \sim 165$ MeV\cite{Junnarkar:2018twb,Leskovec:2019ioa,Mohanta:2020eed}.

Aims of our study are to confirm an existence of a bound $T_{bb}$ and to investigate its properties, 
employing  the HAL QCD potential method\cite{Ishii:2006ec,Aoki:2009ji,Aoki:2012tk}, which is different from methods in previous studies. 
In particular,  we have performed a coupled channel analysis with moving $b$ quarks by NRQCD,
in order to see how competing effects,  the reduction  by the coupled channel and the enhancement by the moving $b$ quarks,
finally determines the binding energy of $T_{bb}$.

\section{Methodology}
\subsection{HAL QCD method}
Since a threshold of the ${\cal B}^*:= \bar B^*\bar B^*$ channel is only about 45 MeV above the ${\cal B}:=\bar B\bar B^*$ threshold,
we have carried out a coupled channel analysis in our study.
Namely, we employ the time-dependent coupled channel HAL QCD method\cite{Ishii:2012ssm,Aoki:2011gt} at the leading order (LO) in the derivative expansion, which is summarized as follows.
The time-dependent $2\times 2$ coupled channel equation reads
\beqa
\underbrace{\left( {\nabla\over 2\mu_\alpha} -{\partial\over \partial t} +{1+\delta_\alpha^2\over 8\mu_\alpha}{\partial^2\over \partial t^2}\right) R^\alpha{}_\xi({\bf r},t)}_{:={\cal K}^\alpha{}_\xi}
\simeq \sum_\beta \tilde\Delta^{\alpha\beta}(t) \int d^3r^\prime U^{\alpha\beta}({\bf r}, {\bf r}^\prime) R^\beta{}_\xi({\bf r}^\prime,t)
\eeqa
where $\alpha,\beta= 0({\cal B}), 1({\cal B}^*)$, $\mu_\alpha$ is a reduced mass of the channel $\alpha$, and
\beqa
\delta_\alpha &:=& {\vert m_{\alpha_1} -m_{\alpha_2}\vert \over m_{\alpha_1} +m_{\alpha_2}}, \quad
\tilde\Delta^{\alpha\beta}(t)=\sqrt{Z_{\beta_1}Z_{\beta_2}\over Z_{\alpha_1}Z_{\alpha_2}}{ e^{-(m_{\beta_1}+m_{\beta_2})t}\over e^{-(m_{\alpha_1}+m_{\alpha_2})t}},
\eeqa
 with $m_{\alpha_i}$ and $Z_{\alpha_i}$ being mass and $Z$-factor of the $i$-th particle in the channel $\alpha$.
Here the normalized correlation function for the source operator ${\cal J}^\dagger_\xi$ with $\xi =0,1$ and hadron operators $H_{\alpha_i}$ is given by
\beqa
R^\alpha{}_\xi ({\bf r},t-t_0) &=& {1\over e^{-(m_{\alpha_1}+m_{\alpha_2})(t-t_0)}}\sum_{\bf x}
\langle 0\vert H_{\alpha_1}({\bf x}+{\bf r},t) H_{\alpha_2}({\bf x},t) {\cal J}^\dagger_\xi (t_0)\vert 0\rangle.
\eeqa 
 A potential matrix at the LO in the derivative expansion of  $U^{\alpha\beta}({\bf r},{\bf r}^\prime)=V^{\alpha\beta}({\bf r})\delta^{(3)}({\bf r} -{\bf r}^\prime)$ is extracted from the above equation as
 \beqa
 \left(
\begin{array}{cc}
V^{00} ({\bf r}) &  \tilde\Delta^{(01)}(t) V^{01} ({\bf r})    \\
  \tilde\Delta^{(10)}(t) V^{10} ({\bf r}) &  V^{11} ({\bf r})\\ 
\end{array}
 \right) &=&
  \left(
\begin{array}{cc}
{\cal K}^0{}_0 ({\bf r}) &   {\cal K}^0{}_1 ({\bf r})   \\
{\cal K}^1{}_0 ({\bf r}) &   {\cal K}^1{}_1 ({\bf r})   \\
\end{array}
 \right)
   \left(
\begin{array}{cc}
R^0{}_0 ({\bf r}) &  R^0{}_1 ({\bf r})   \\
R^1{}_0 ({\bf r}) &  R^1{}_1 ({\bf r})   \\
\end{array}
 \right)^{-1} .
 \eeqa
 A mild $t$ dependence of  the LO potential matrix extracted in the above procedure  indicates that
 inelastic contributions as well as truncation errors in the LO approximation are small.
  
\subsection{NRQCD action for heavy quarks}
Since $b$ quarks are too heavy to treat relativistically at lattice spacings in currently available gauge configurations,
we have employed the NonRelativistic QCD (NRQCD) formulation\cite{Thacker:1990bm} for $b$ quarks, where
a time evolution of  the two-spinor NRQCD propagator $G_\psi$ is controlled by the Hamiltonian ${\cal H}_\psi:= {\cal H}_0+\delta{\cal H}$ as
\beqa
G({\bf x}, t+1\vert s_0) &=& \left(1-{{\cal H}_0\over 2n}\right)^n \left( 1-{\delta{\cal H}\over 2}\right) U_4^\dagger(x) \left( 1-{\delta{\cal H}\over 2}\right)
 \left(1-{{\cal H}_0\over 2n}\right)^n G({\bf x}, t\vert s_0)+s_0({\bf x)}\delta_{t,-1}, ~~~~
\eeqa
where ${\cal H}_0= -{1\over 2M} \Delta^{(2)}$ is the leading Hamiltonian  at $O(v^2)$ with $M$ and $v$ being a mass and a velocity of a $b$ quark, respectively,  $s_0$ is a source vector, and $n=2$ is a stabilized parameter.
In our study, we take $O(v^4)$ terms for $\delta {\cal H}= \sum_{i=1}^6 c_i \delta {\cal H}^{(i)}$ with
\beqa
\delta{\cal H}^{(1)} &=& -{1\over 2M} \sigma\cdot {\bf B}, \
\delta{\cal H}^{(2)} ={i\over 8M^2}\left(\nabla\cdot {\bf E} -{\bf E}\cdot\nabla\right),\ 
\delta{\cal H}^{(3)} =-{1\over 8M}\sigma\cdot \left(\nabla\times {\bf E} -{\bf E}\times\nabla\right),\nn\\
\delta{\cal H}^{(4)} &=& -{1\over 8M^3} \left(\Delta^{(2)}\right)^2, \
\delta{\cal H}^{(5)} ={1\over 24M}\Delta^{(4)},\ 
\delta{\cal H}^{(6)} =-{1\over 16nM^2}  \left(\Delta^{(2)}\right)^2,
\eeqa
where  all $c_i$'s are taken to be tree level values ($c_i=1$) with the tadpole improvement that $U_\mu\to U_\mu/u_0$\cite{Lepage:1992tx},
$\nabla$, $\Delta^{(2)}$, $\Delta^{(2)}$ are 1st, 2nd and 4th order symmetric covariant difference in space, respectively, and the chromo-electromagnetic fields ${\bf E}$ and ${\bf B}$ are given by the standard clover-leaf definition.
Correspondingly, the FWT transformation matrix at $O(v^4)$\cite{Ishikawa:1997sx} is employed in our study.

A mass of heavy-light meson $X$ with the NRQCD $b$ quark is extracted as
\beqa
M_X &=& {{\bf p}^2 -\left(E_X({\bf p})-E_X({\bf 0}) \right)^2\over 2 \left(E_X({\bf p})-E_X({\bf 0}) \right)},
\eeqa
where $E_X({\bf p}) =\sqrt{{\bf p}^2 + M_X^2} -\delta$ is an energy of the $X$ meson with a momentum ${\bf p}$, and
$\delta$  is an additive mass renormalization for the $b$ quark. We do not have to know $\delta$ in the above formula. 

\subsection{Operators}
For the coupled channel analysis($\alpha=0,1$), we employ 2 types of meson-meson local sink operators given by
\beqa
{\cal B}_j({\bf r}) &:=& \sum_{\bf x}\left\{ B_{u}({\bf x}+{\bf r}) B_{d,j}^*({\bf x}) - [u\leftrightarrow d]\right\},\
B_{q}({\bf x}):=\bar q({\bf x})\gamma_5 b({\bf x}),\   B_{q,j}^*({\bf x}):=\bar q({\bf x})\gamma_j b({\bf x}),\nn  \\
{\cal B}^*_j({\bf r}) &:=& \epsilon_{jk\ell}\sum_{\bf x}\left\{ B^*_{u,k}({\bf x}+{\bf r}) B_{d,\ell}^*({\bf x}) - [u\leftrightarrow d]\right\}.
\eeqa   
Since both ${\cal B}_j^\dagger$ and $({\cal B}_j^*)^\dagger$ source operators create similar combinations of two independent states,
we introduce a diquark operator made of
heavy and light diquarks as
\beqa
{\cal J}_{{\cal D}_j^\dagger} &=& \left(\epsilon^{abc} \bar b^b \gamma_j C \bar b^c\right)  
 \left(\epsilon^{ade} \bar d^d C \gamma_5 \bar u^e\right) -\left[ u\leftrightarrow d\right],
\eeqa
where $C=\gamma_4\gamma_2$ is a charge-conjugation matrix.
To create two different $ R^\alpha{}_\xi$'s  ($\xi=0,1$)  with wall quark sources in  the Coulomb gauge fixing,
we take a meson-meson operator ${\cal J}_{{\cal B}_j^\dagger}$ made of the creation operator of ${\cal B}_j$  for $\xi=0$ and a diquark operator ${\cal J}_{{\cal D}_j^\dagger} $ for $\xi=1$.

\subsection{Lattice QCD configurations}
In our study, we have employed $2+1$ flavor full QCD configurations generated by the CP-PACS Collaboration\cite{PACS-CS:2008bkb} with the Iwasaki gauge action and the Wilson-Clover light quark action at $a\simeq 0.09$ fm, and 
400 configurations are used  at each pion mass, $m_\pi = 701, 571,461$ MeV.
We tune the $b$ quark mass $M_b$ to reproduce the spin-averaged $b\bar b$ mass, $M_{b\bar b}^{\rm spin-avg.}\simeq 9450$ MeV,
leading to $M_{\bar B}^{\rm spin-avg}:=(M_{\bar B} +3 M_{B^*})/4 =5440(174)$, 5382(269), and 5332(220) MeV at each pion mass, which agree with an experimental value, 5313 MeV within 5\% statistical errors, while the spin-splitting  $\Delta E_{\bar B\bar B^*}:= M_{\bar B^*} -  M_{\bar B}$ decreases as 49.4(2.6), 44.9(1.6) and 42.7(3.9) MeV.

\section{Numerical results}
\subsection{Leading order potentials}
\begin{figure}[htb]
\centering
\vskip -2.5cm
  \includegraphics[angle=0, width=0.9\textwidth]{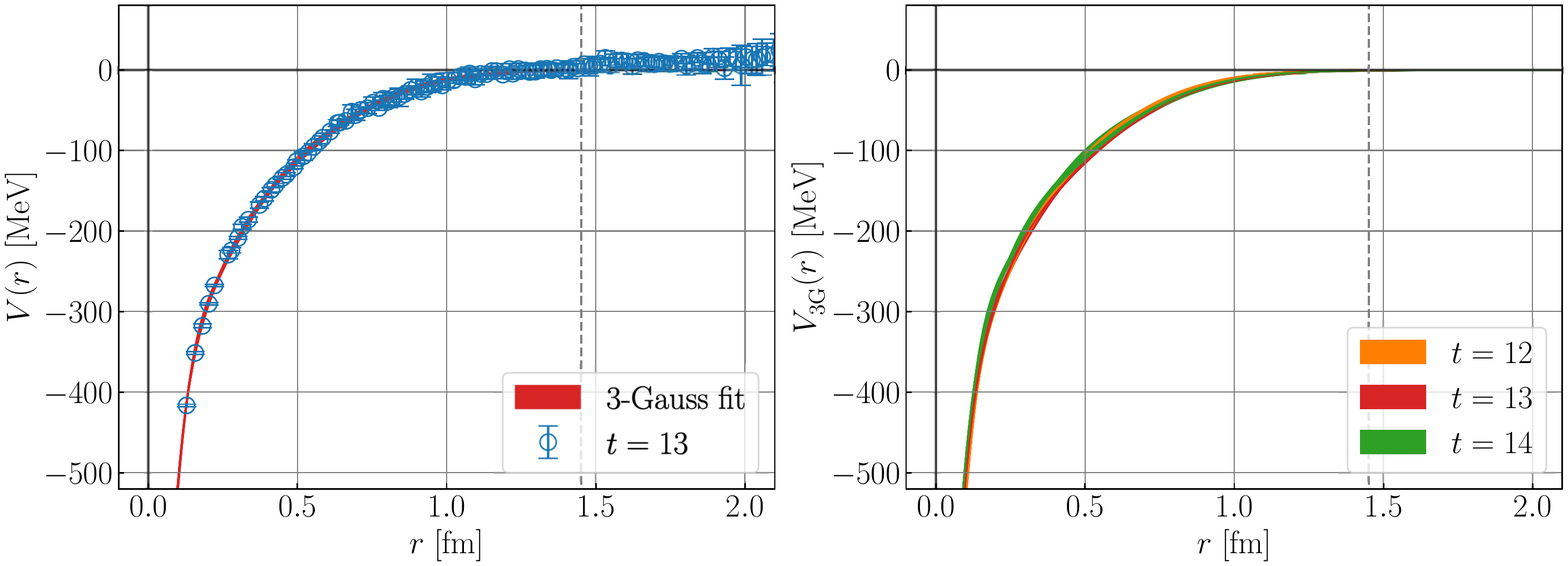}
 \vskip -2.5cm
 \caption{(Left) A single channel potential (blue circles) extracted at $t=13$, together with the 3 Gaussian fit by a red line.
 (Right) A comparison among fits at $t=12,13,14$.
 }
 \label{fig:pot_single}
\end{figure}
Fig.~\ref{fig:pot_single} (Left) shows that a potential for a single ${\cal B}$ channel is attractive at all distances smaller than 1.0 fm  and 
is well described by a sum of 3 Gaussians (red line).
A small $t$ dependence of the potential in  Fig.~\ref{fig:pot_single} (Right) suggests
that contributions from inelastic states as well as higher order terms in the derivative expansion are well under control.

\begin{figure}[bht]
\centering
  \includegraphics[angle=0, width=0.85\textwidth]{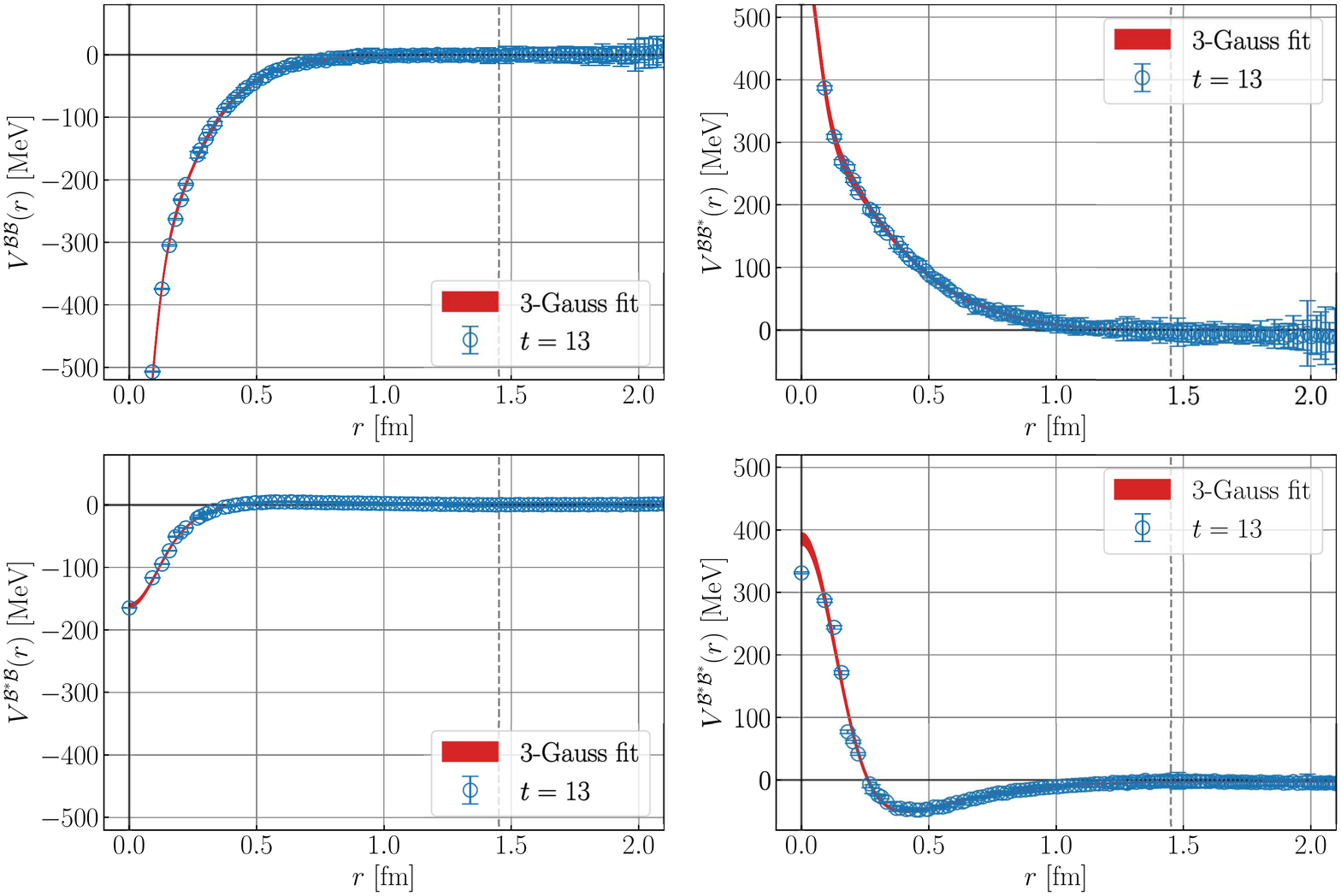}
 \caption{ $2\times 2$ coupled-channel potentials (blue circles) extracted at $t=13$, together with 3 Gaussian fits by red lines.
 }
 \label{fig:pot2x2}
\end{figure}
Fig.~\ref{fig:pot2x2} shows a LO coupled channel potential at $m_\pi= 701$ MeV, extracted at $t=13$.
A diagonal potential $V^{{\cal B}{\cal B}}$ is attractive at all distances smaller than 0.8 fm but the range of attraction is shorter than 1.0 fm in the single channel, while $V^{{\cal B}^*{\cal B}^*}$ has a repulsive core surrounded by an attractive pocket at $r\simeq 0.4$ fm. 
We have surprisingly found that off-diagonal parts are not symmetric leading to a large violation of the Hermiticity.
In addition, off-diagonal interaction between ${\cal B}$ and ${\cal B}^*$ are comparable to diagonal ones in magnitudes.
Therefore, such strong off-diagonal interactions should be taken into account in the scattering analysis even for the single channel scattering
below the ${\cal B}^*$ threshold, as will be seen in the next subsection.

The coupled channel potential is also well fitted by a sum of 3 Gaussian functions (red lines in Fig.~\ref{fig:pot2x2}) at each pion mass.
As the pion mass decreases, both diagonal and off-diagonal potentials become stronger and more long-ranged.
Thus a mixing effect due to off-diagonal potentials remains relevant even at the physical pion mass.

\subsection{Scattering analysis}
Since the Hermiticity of the coupled channel potential is badly violated, probably due to the LO approximation, we consider a single channel scattering
in the ${\cal B}$ channel below the ${\cal B}^*$ threshold while employing $2\times 2$ coupled channel potential $V^{XY}$ to include virtual ${\cal B}^*$ states.
Integrating out virtual ${\cal B}^*$ contributions, an effective single channel potential becomes non-local and energy-dependent as
\beqa
U^{{\cal B}{\cal B}}_{{\rm eff},E}({\bf x},{\bf y}) = V^{{\cal B}{\cal B}}({\bf x})\delta^{(3)}({\bf x}-{\bf y})
+ V^{{\cal B}{\cal B}^*}({\bf x}) G_E^{{\cal B}^*{\cal B}^*}({\bf x},{\bf y})  V^{{\cal B}^*{\cal B}}({\bf y}) ,
\eeqa
where $G_E^{{\cal B}^*{\cal B}^*}$ is a propagator of ${\cal B}^*$ in the presence of the diagonal potential $V^{{\cal B}^*{\cal B}^*}$ at energy $E$.

\begin{figure}[bth]
\centering
  \includegraphics[angle=0, width=0.90\textwidth]{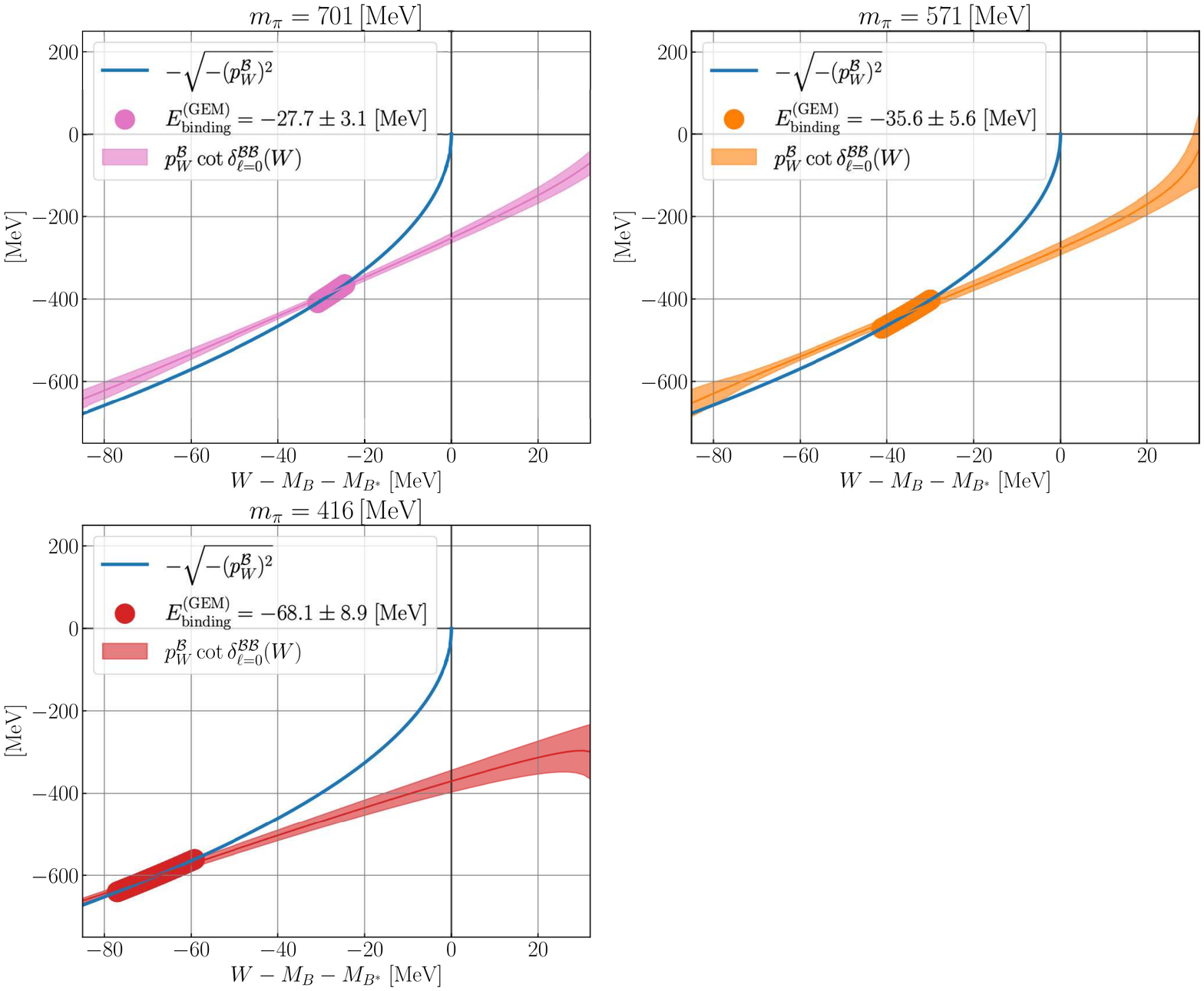}
 \caption{ Results of $p\cot\delta(p)$ in the ${\cal B}$ channel as a function of the  energy measured from the ${\cal B}$ threshold, $W-M_{\bar B} -M_{\bar B^*}$,
  at $m_\pi=701$ (pink), 571 (orange), and 416 (red) MeV, 
  together with the bound state condition, $-\sqrt{-p^2}$, by a blue solid curve.
  A think band along $-\sqrt{-p^2}$ curve shows a binding energy calculated by the GEM, which agrees well with 
  an intersection between the  $p\cot\delta(W)$ and the bound state condition.
 }
 \label{fig:pcotd_mpi}
\end{figure}
Solving the Lippmann-Schwinger equation with the fitted coupled channel potentials, we extract scattering phase shift $\delta(p)$ in the single channel ${\cal B}$ below the inelastic  threshold of the channel ${\cal B}^*$, where $p$ is a magnitude of momentum extracted from the center of mass energy as
$W=\sqrt{p^2+M_{\bar B}^2}+\sqrt{p^2+ M_{\bar B^*}^2}$.
 Fig.~\ref{fig:pcotd_mpi} shows $p\cot\delta(p)$ as a function of the energy from the ${\cal B}$ threshold, $W-M_{\bar B} - M_{\bar B^*}$, at $m_\pi=701$ (pink), 571 (orange), and 416 (red) MeV.
 At $0 < W - M_{\bar B} - M_{\bar B^*}< M_{\bar B^*}-M_{\bar B}\simeq 45$ MeV, phase shifts $\delta(p)$ is physical, while
 an existence of a bound state is examined by its analytic continuation  at $W - M_{\bar B} - M_{\bar B^*} < 0$.
 Since an analytic continuation of the on-shell $T$-matrix has a pole at $p\cot \delta(p) = i p$, an intersection between $p\cot\delta(p)$ (pink, orange and  red bands) and $-\sqrt{-p^2}$ in Fig.~\ref{fig:pcotd_mpi} corresponds to a bound states in the ${\cal B}$ channel, showing that there exists one bound $T_{bb}$ state  at each pion mass. Note that the intersection at each pion mass satisfies the physical pole condition\cite{Iritani:2017rlk} that
 \beqa
 \left. {d^2 \over dp^2} \left[ p\cot\delta(p) - (-\sqrt{-p^2})\right] \right\vert_{p^2= -p^2_{\rm BS}} < 0,
 \eeqa
 where $p_{\rm BS}$ corresponds to a magnitude of the bound state momentum.
 We also calculate the binding energy solving the Schr\"odinger equation directly by the Gaussian Expansion Method (GEM)\cite{Kamimura:1988zz},
 and results represented by thick lines along the bound state condition (blue curve)  agree with ones by $p\cot\delta(p)$, as seen in Fig.~\ref{fig:pcotd_mpi}. 
 
\begin{figure}[bth]
\centering
  \includegraphics[angle=0, width=0.48\textwidth]{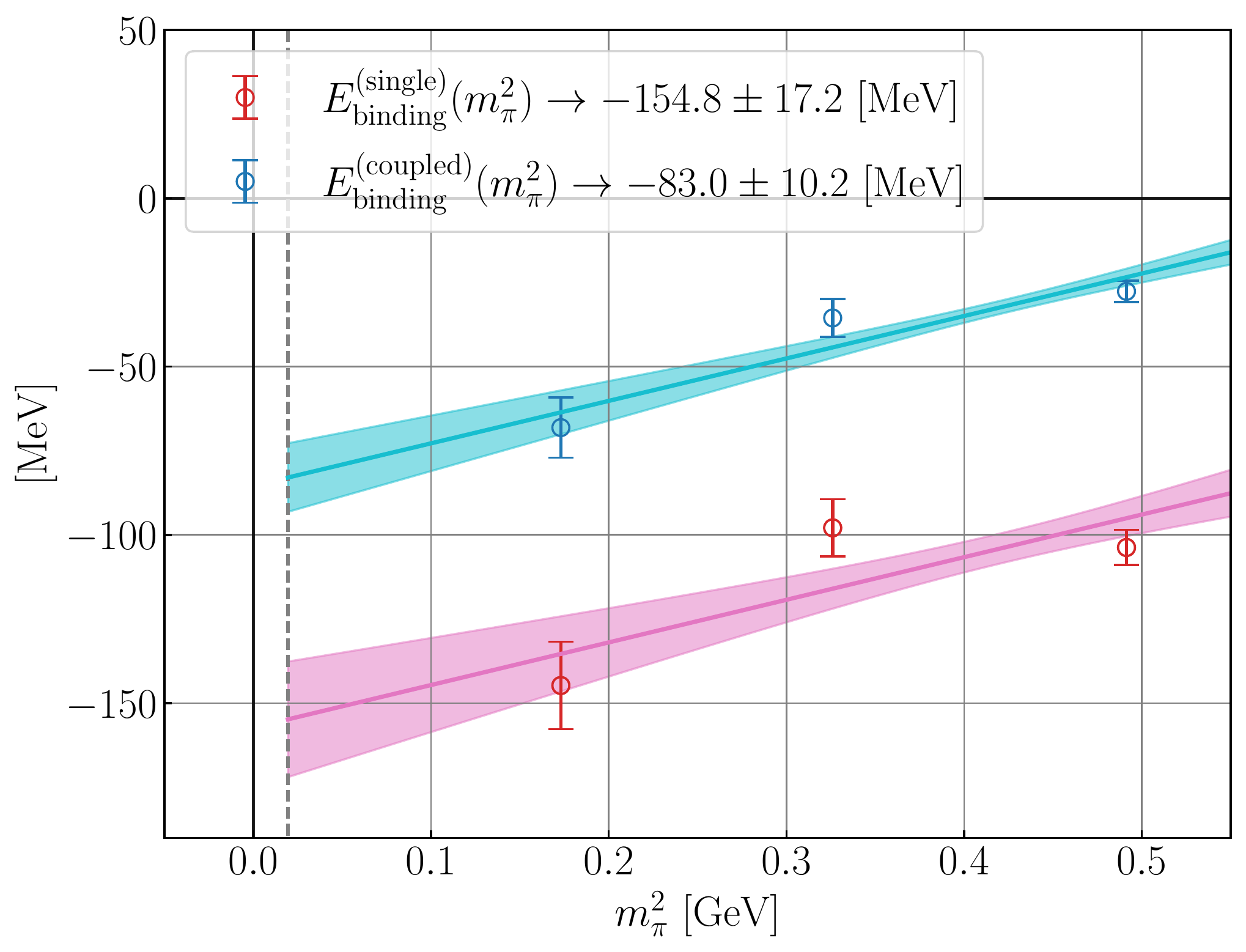}
  \includegraphics[angle=0, width=0.48\textwidth]{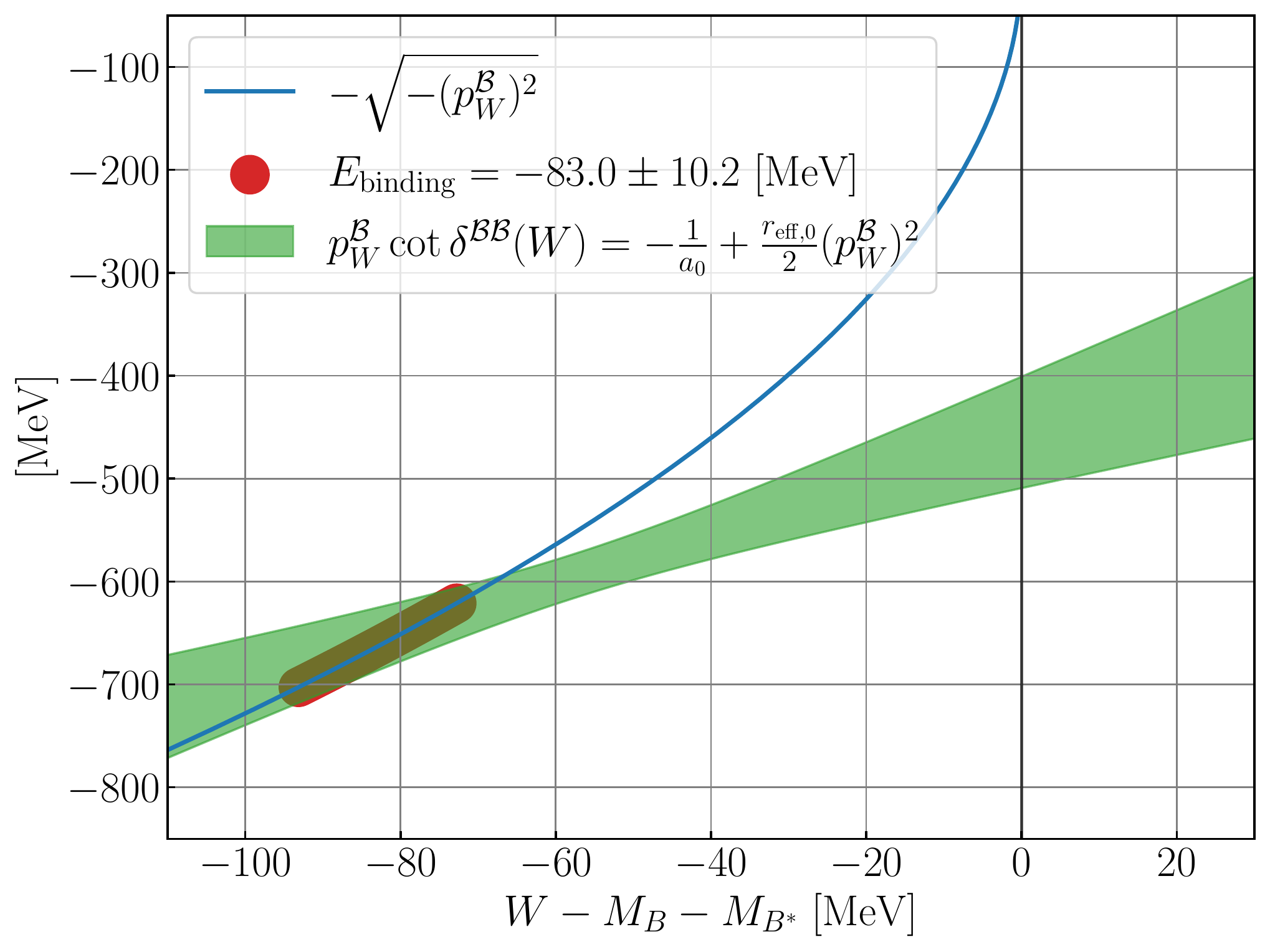} 
 \caption{(Left) The binding energy obtained by the GEM as a function of the pion mass squared $m_\pi^2$, together with a linear chiral extrapolation in $m_\pi^2$ to $m_\pi = 140$ MeV by solid lines. Results from the single channel analysis (magenta circles) and the coupled channel analysis (cyan circles) are shown. \\
 (Right) The ERE at $m_\pi=140$ MeV (green band), obtained with $a_0$ and $r_{{\rm eff},0}$ by linear extrapolations in $m_\pi^2$, together with the bound state condition, $-\sqrt{-p^2}$ (blue solid curve). An intersection between the two satisfies the physical pole condition and, moreover, agrees well
 with the binding energy by the GEM at $m_\pi=140$ MeV (red thick curve along $-\sqrt{-p^2}$).
 }
 \label{fig:pot_comp}
\end{figure}
Fig.~\ref{fig:pot_comp} (Left) shows the binding energy by the GEM as a function of $m_\pi^2$ from the coupled channel analysis (cyan circles),
together with a chiral extrapolation linear in $m_\pi^2$ by a blue solid. For a comparison, the result from the standard single channel analysis, where
the potential in Fig.~\ref{fig:pot_single} is used, for example,
  is also given by magenta circles. At the physical pion mass, $m_\pi = 140$ MeV, we obtain
\beqa
E_{\rm binding}^{\rm (single, phys)} &=& -154.8\pm 17.2\ \mbox{MeV}, \quad
E_{\rm binding}^{\rm (coupled, phys)} = -83.0\pm 10.2\ \mbox{MeV},
\label{eq:BE}
\eeqa
which shows roughly a 50 \% reduction of the binding energy due to contributions from virtual  ${\cal B}^*$ states.
Thus the coupled channel analysis is indeed important to estimate the binding energy for $T_{bb}$ more precisely. 

As a cross check, we have performed the chiral extrapolation of the effective range expansion (ERE) parameters $a_0$ and $r_{{\rm eff},0}$,  obtained from a linear fit in $p^2$ as $p\cot\delta(p) = -{1\over a_0} + {r_{{\rm eff},0}\over 2} p^2$ at each pion mass. A linear extrapolation in $m_\pi^2$ leads to
$a_0 =0.43(5)$ fm and $r_{{\rm eff},0}=0.18(6)$ fm at $m_\pi=140$ MeV in the coupled channel analysis.
Then the binding energy at the physical point ($m_\pi=140$ MeV) is alternatively estimated from an intersection between the ERE  with these $a_0$ and $r_{{\rm eff},0}$ (green band) and the bound state condition $-\sqrt{-p^2}$ (blue solid curve), as shown in Fig.~\ref{fig:pot_comp} (Right). The intersection satisfies the physical pole condition and agrees well with the binding energy by the GEM directly extrapolated to the physical point.  
This agreement in the binding energy between two extrapolations to the physical point provides a validity of our analysis. 

\section{Conclusion}
\begin{figure}[bth]
\centering
\vskip -0.5cm
  \includegraphics[angle=0, width=0.75\textwidth]{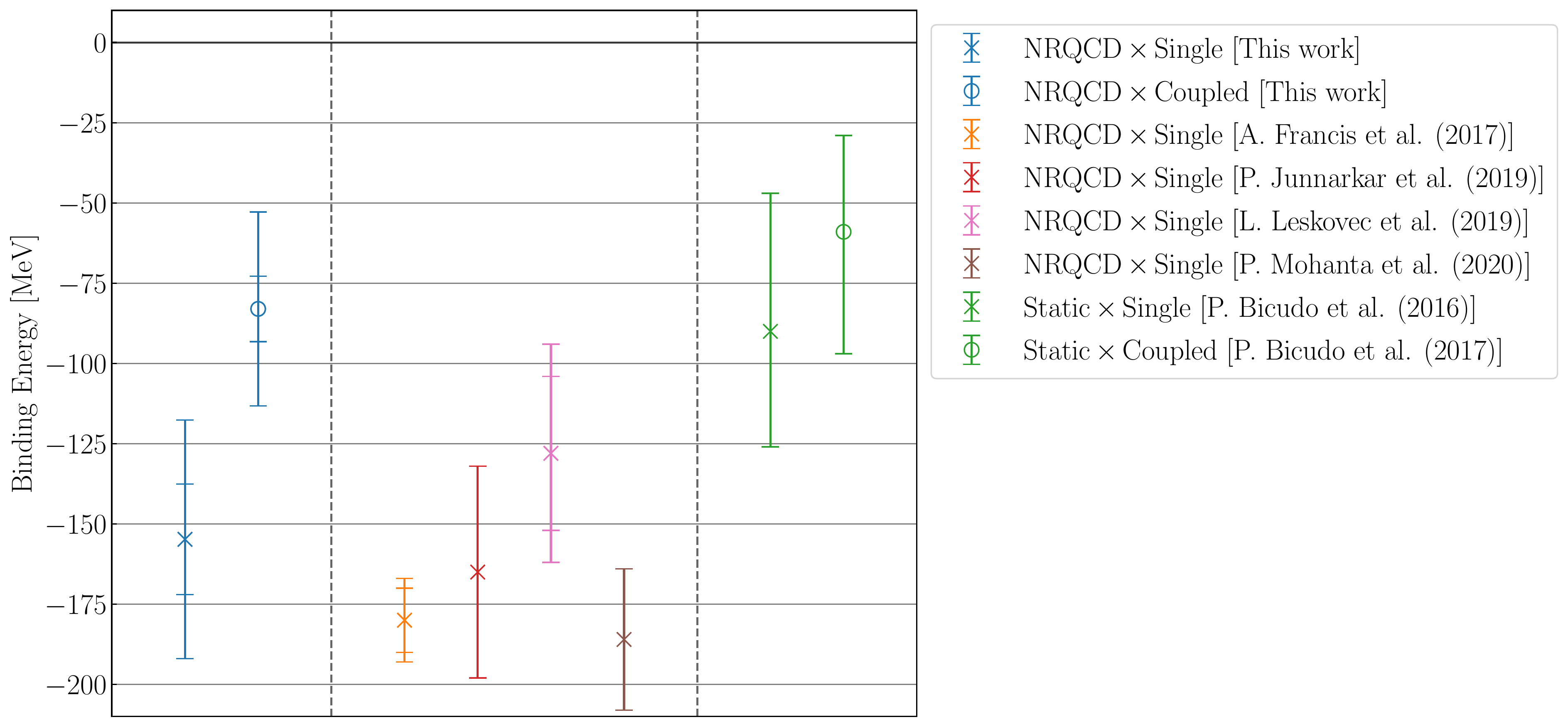}
 \caption{A comparison of binding energies for the tetra uark bound state $T_{bb}$ among several lattice QCD calculations,
 our results (blue), spectra with the NRQCD (orange, red, magenta and brown), and the static quark potential (green),
 from single channel (crosses) and couple channel (circles) analyses.
 }
 \label{fig:pot_comp_all}
\end{figure}
We have calculated $S$-wave channel potentials between  $\bar B$ and $\bar B^*$, and  found one bound $T_{bb}$ state at three pion masses.
The linear chiral extrapolation of the binding energy in $m_\pi^2$ gives  \eqref{eq:BE},
and adding an empirical systematic error of $\pm 20$ MeV to each value, we compare our results with other lattice results in Fig.~\ref{fig:pot_comp_all}.
First of all we observe a consistency in the binding energy of $T_{bb}$ among the single channel analysis with the NRQCD for $b$ quarks including our results. 
Secondly  the biding energy of $T_{bb}$ increases if the treatment of the $b$ quark on the lattice is changed from the static quark to the NRQCD.
Thirdly an inclusion of virtual ${\cal B}^*$ effects reduce the binding energy of $T_{bb}$. In particular, the reduction becomes 50\% using 
the HAL QCD potential combined with the NRQCD for $b$ quarks.\\

\noindent
Our numerical calculation has been performed on Yukawa-21 at YITP, Kyoto University. This work is supported in part by the JSPS Grant-in-Aid for Scientific Research (Nos.~JP16H03978, JP18H05236).

\end{document}